\documentclass{aa}  

\usepackage{graphicx}
\usepackage{txfonts}
\usepackage{natbib} 
\bibpunct{(}{)}{;}{a}{}{,} 

\usepackage{tikz}
\usetikzlibrary{positioning}
\usepackage{hyperref}   
\hypersetup{colorlinks=true,linkcolor=blue,citecolor=blue,filecolor=blue,urlcolor=blue}

\begin{document}

   \title{Robust statistical tools for identifying multiple stellar populations in globular clusters in the presence of measurement errors.} 

   \subtitle{A case study: NGC 2808}

   \author{G. Valle \inst{1}, M. Dell'Omodarme \inst{1}, E. Tognelli \inst{2}
          }
   \titlerunning{Statistical tools for MP detection in globular clusters}
   \authorrunning{Valle, G. et al.}

   \institute{
        Dipartimento di Fisica "Enrico Fermi'',
        Universit\`a di Pisa, Largo Pontecorvo 3, I-56127, Pisa, Italy
 \and
        CEICO, Institute of Physics of the Czech Academy of Sciences, Na Slovance 2, 182 21 Praha 8, Czechia}    

   \offprints{G. Valle, valle@df.unipi.it}

   \date{Received ; accepted }

  \abstract
{The finding of multiple stellar populations (MPs), which are defined by patterns in the stellar element abundances, is considered today a distinctive feature of globular clusters. However, while data availability and quality have improved in the past decades, this is not always true for the techniques that are adopted to analyse them, which creates problems of objectivity for the claims and reproducibility. }
{Using NGC 2808 as test case, we show the use of well-established statistical clustering methods. We focus our analysis on the red giant branch phase, where two data sets are available in the recent literature for low- and high-resolution spectroscopy.}
{ We adopted hierarchical clustering and partition methods. We explicitly addressed the usually neglected problem of measurement errors, for which we relied on techniques that were recently introduced in the statistical literature. The results of the clustering algorithms were subjected to a silhouette width analysis to compare the performance of the split into different numbers of MPs. }
{For both data sets the results of the statistical pipeline are at odds with those reported in the literature. Two MPs are detected for both data sets, while the literature reports five and four MPs from high- and low-resolution spectroscopy, respectively. The silhouette analysis suggests that the population substructure is reliable for high-resolution spectroscopy data, while the actual existence of MP is questionable for the low-resolution spectroscopy data.
The discrepancy with literature claims can be explained with the different methods that were adopted to characterise MPs. By means of Monte Carlo simulations and multimodality statistical tests, we show that the often adopted study of the histogram of the differences in some key elements is prone to multiple false-positive findings. }
{The adoption of statistically grounded methods, which adopt all the available information to split the data into subsets and explicitly address the problem of data uncertainty, is of paramount importance to present more robust and reproducible research.} 


   \maketitle

\section{Introduction}\label{sec:intro}

The presence of multiple stellar populations (MPs) is considered today a distinctive feature of almost all globular clusters (GCs). Despite important theoretical efforts, such as the recent suggestion of the role of stellar mergers in the MP origin \citep{Wang2020}, a clear understanding of the MP formation is still lacking \citep[see e.g.][and references therein]{Bastian2018}. A typical feature of these  MPs is an abundance variation in light elements with little or no dispersion in heavy elements, with significant element abundance anti-correlations. 
A widely used signature to identify MPs inside a GC is the distribution of key abundance indexes such as [O/Na] or CN-CH \citep[e.g.][]{Carretta2009b, Carretta2015, Hong2021}. The MP identification is often based on judgments made by naked eye, guided by of apparent splits in the colour-magnitude diagram (CMD) or peaks in the element abundances. In these cases, the classification of a star into a given group, and the number of identified populations itself, is based on subjective judgement and not on statistical methods.
An obvious drawback of this approach is the lack of robustness of the results and severe reproducibility issues. In fact, different researchers may get different data groupings, without the possibility of comparing these alternative claims.   
 
Moreover, basing the grouping definition on the analysis of patterns in the abundance versus abundance index plots, as is frequently done, is risky because of how individuals perceive  visual representations.  
Gestalt principles of perceptual organisation \citep{Wertheimer1938} emphasize that organisms perceive entire patterns or configurations and not the individual components. The fundamental law that governs a Gestalt principle is that we tend to order our experience in a regular, orderly, and recognizable manner.
The Gestalt laws of proximity, similarity, connectedness, continuity, and common fate determine how individual graphical features are grouped together to form coherent representations \citep{Ali2013, Pinker1990}.
In particular, the law of proximity states that when an individual perceives an assortment of objects, they perceive objects that are close to each other as forming a group.
These factors can lead to an overestimation of subgroups in a population because random gaps are perceived as representing a real boundary between subpopulations.

Fortunately, there are a number of statistically validated techniques to deal with the grouping problem. Even when robust statistics-based clustering methods are adopted to identify MPs \citep[e.g.][]{Carretta2011, Gratton2011, Simpson2012, pasquato2019multiple}, however, several paramount aspects are usually not addressed and therefore deserve particular attention.   
The first aspect is the impact of the data uncertainties. This is a relevant issue because chemical abundances are affected by non-negligible uncertainties that are often comparable with the intercluster variability.
The second problem is that statistical clustering methods do not directly provide information about the optimal grouping, that is, whether a two-group split is better than a three-group split. While several statistical methods exist to investigate this, they are usually ignored. 

The aim of this paper is to illustrate the strength of some methods, using as a case study one of the most frequently studied GCs, that is, NCG 2808. This cluster is a perfect target for this investigation because multiple stellar populations have been reported in the literature in different evolutionary stages, from the main sequence to the asymptotic giant branch \citep[see e.g.][]{DAntona2005, Piotto2007, DAntona2008, Lee2009, Dalessandro2011, Carretta2015, Marino2017}.

The paper is organised as follows. Section~\ref{sec:method} introduces the adopted statistical methods and the data sets we used in the analysis. Section~\ref{sec:results} contains the results of the cluster analysis for the two data sets. Some concluding remarks are collected in Section~\ref{sec:conclusions}. The paper contains a statistical appendix in which the adopted statistical methods are presented in some detail.

\section{Methods and data}\label{sec:method}

\subsection{Statistical methods}

Identifying multiple stellar populations in GCs by their chemical abundances is a classical problem that is addressed in statistic literature in the vast area of unsupervised clustering techniques. These methods attempt to split data into groups that are inferred from data patterns, so that the intergroup difference is as small as possible, while the intragroup difference is as large as possible.

These unsupervised techniques can be broadly divided into hierarchical clustering (HC) methods and partition methods. Partition methods may additionally be split into distance-based, density-based, and model-based methods \citep[see e.g.][]{KaufmanRousseeuw90, venables2002modern, simar, Feigelson2012,SAXENA2017}. While these methods were not developed to take measurement errors into account,  they have recently been further generalised to cover this important topic \citep[e.g.][]{Kumar2007, Su2018, pankowska2020effect}.  

In the following we adopt three different clustering methods: a divisive HC, an agglomerative HC, and a method called partition around medoids (PAM). The rationale for using more than one clustering algorithm is that none of them is the absolute best for any data set; indeed they all have known weaknesses, and comparing their performances usually helps to better understand the underlying data structure.  For each method, we tested different possible data splits, from two to five groups, then we evaluated the average  silhouette, which is an indicator of how clearly the elements belong to a given group or cluster. An overview of the adopted methods is given in Appendix~\ref{sec:app-tools}. The analysis was conducted using R 4.1.0 \citep{R}\footnote{The computational methods adopted in the paper are demonstrated at {\url{https://github.com/mattdell71/cluster-analysis}} on the low-resolution spectroscopy data set}.

A crucial point in this analysis is the possibility of including  observational errors. We took the errors in the abundances at two different levels into account. First, errors were used to construct the dissimilarity matrix (see Appendix~\ref{sec:app-tools} for details): briefly, the dissimilarity matrix is the metric adopted to divide objects into groups relying on the clustering techniques described above. The objective of these methods is to group objects with low 
dissimilarities. Neglecting the data uncertainties and adopting a straightforward Euclidean distance in the abundances hyperspace would ignore that some elements are known to a better  precision; the same raw difference is therefore quite different according to the considered index. However, this is not enough to fully account for the measurement error problem. A probabilistic clustering is advised in this case. Therefore
we repeated the cluster analysis in the framework proposed by \citet{Su2018}. This approach adopts a Monte Carlo (MC) resampling of the data set to assess the probability that an object might belong to a given cluster (see Appendix~\ref{sec:error-var} for details). According to this analysis, we can only obtain the probability that an object belongs to a given cluster and not a rigid clustering. It also allows us to evaluate how the average silhouette of different grouping responds to the data uncertainty.

A widespread approach to the identification of MPs by chemical abundances or photometric data is the examination of the differences in key elements, such as [O/Na] or CN-CH \citep[e.g.][]{Carretta2009b, Carretta2015, Simpson2017, Guerco2019, Hong2021}, or differences in photometric bands or colour indices \citep[e.g.][]{DAntona2005, Dalessandro2011, Marino2017}. MPs are then identified by comparing the element abundances with those of field stars or by peaks in the histogram of these indexes. This approach does not determine the reliability of the detected substructure, however. Furthermore, two main drawbacks hinder the adoption of peaks in the histogram as tracers of MPs, both related to binning: the bin width, and the starting position of the first bin \citep[see e.g.][]{simar}. A better approach to approximate the empirical probability density function is adopting a kernel density estimator (see Appendix~\ref{sec:kernel} for details). Roughly speaking, a kernel density estimator can be considered a smoothed histogram without the dependence on the starting position. The bin width is replaced by the so-called bandwidth $h$, and several well-tested approaches exist to estimate it. This parameter can be considered as a moving smoothing window. A too large $h$ results in a very smooth density estimator, which may mask interesting features, while a too low $h$ produces a noisy estimator with many spurious peaks. 
General-purpose widespread algorithms exist to estimate $h$ (e.g. the SJ estimate from \citealt{Sheathe1991} and the Wand estimate from \citealt{Wand1994}). In addition to these,
we also adopted a third density estimator, the decon approach by \citet{Wang2011}, which corrects the density estimate considering data uncertainties.

The subjective judgement of multimodality in the empirical density was further investigated by means of two statistical tests: the Dip test, which is a statistical tool for determining a departure for unimodality \citep{Hartigan1985}, and which is implemented in the diptest R library \citep{diptest}, and the \citet{Ameijeiras2019} excess mass test, which is available in the multimode R library \citep{multimode}. 

Finally, to show some issues arising from the adoption of an empirical density estimator as a tool for identifying MPs, we performed some MC simulations (Sect.~\ref{sec:peaks}). 
We simulated random points from a single population in a 2D space, and we compared the number of subpopulations inferred by peak detection and divisive HC.

\subsection{Data}

As a difference from \citet{pasquato2019multiple}, who showed the use of several clustering methods for NCG 2808, we did not make use of the photometric chromosome map \citep[see e.g.][]{Milone2015, Milone2017}, but we adopted element abundances for red giant branch (RBG) stars from \citet{Carretta2015} (high-resolution spectroscopy) and \citet{Hong2021} (low-resolution spectroscopy). However, the methods presented in this paper can be easily extended to also incorporate pseudo-colours in the analysis.

Abundances of Na, O, Mg, and Si from high-resolution spectroscopy from \citet{Carretta2015} were used, as presented in their Table~4. To perform the cluster analysis, objects with missing values in any of these elements were excluded. This resulted in a sample of 116 RGB stars. Data preprocessing was needed because some element abundances did not have an accompanying error. In these cases, the missing values were replaced by the mean of the error of the specific element over the whole data set.
The adoption of this flat average error is justified by the fact that the uncertainties in the selected quantities are not correlated with the element abundances, as verified with multivariate linear regressions.

Data from low-resolution spectroscopy were obtained from \citet{Hong2021}, who      
measured spectral indexes for CN, CH, and Ca II H\&K lines as tracers of N, C, and
Ca abundances, respectively, from a sample of 91 RGB stars. To remove the effects of gravity and temperature from these indexes, the authors detrended them by subtracting the mean value of the regression against the magnitude V over the whole sample, obtaining the $\delta$CN, $\delta$CH, and $\delta$HK' indexes, which we adopt in the present analysis.

\section{Results}\label{sec:results}

The clustering we evaluated was obtained using two data sets: 1) the indexes $\delta$CN, $\delta$CH, and $\delta$HK' obtained from low-resolution spectra, and 2) the abundances of Na, O, Mg, and Si estimated using high-resolution spectra. The analysis was performed in the same way for both data sets, and the corresponding results are discussed in detail in Sect.~\ref{sec:res-low} and Sect.~\ref{sec:res-high}.

As a preliminary step, we explored the data clustering by means of divisive HC, agglomerative HC, and PAM methods. The functions {\it diana}, {\it agnes}, and {\it pam} in the R library cluster \citep{cluster2021} were used for the computations. We verified that the comparisons of the silhouette values did not show relevant differences among the three adopted algorithms. In the following, we therefore only present the results obtained from divisive HC, which provided the strongest clustering in all the analysed cases. We accordingly adopted the average silhouette from divisive HC to evaluate alternative splits throughout the paper.

The capability of simultaneously taking all the available data into account is another aspect that makes these methods superior to the splitting that is obtained by only considering one index and using other elements to validate the grouping. However, before a multivariate data set is used in the analysis, the variability range or scale of the employed data has to be checked.  When variables vary in very different intervals or ranges, it is advised to rescale them to a new data set so that every variable has zero mean and unit variance \citep[e.g.][]{Rousseeuw1987}. If the mean value of a variable is much higher than the others (which can be the case when abundances and magnitudes or effective temperatures are analysed simultaneously), it dominates the analysis and masks the effect of all the others.
In our case, there was no need to scale data to zero mean and unit variance before clustering. However, we also verified that scaling the data as well as the errors before the clustering did not modify the results. 

\begin{figure*}
        \centering
        \includegraphics[height=16.0cm,angle=-90]{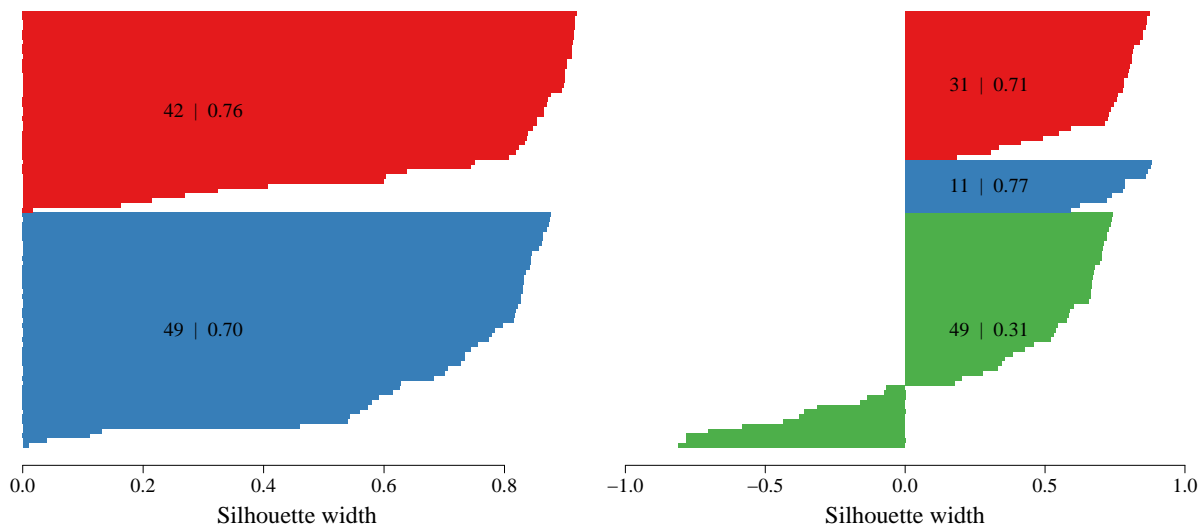}
        \caption{{\it Left}: Silhouette plot of the divisive HC applied to the low-resolution spectroscopy data set for a two-group split. {\it Right}: Same as in the left panel, but for a three-group split. The bars correspond to the silhouette values of objects clustered in the first group (red), in the second group (blue), and in the third group (green). The average silhouettes for the groups are listed along with the group sizes.}
        \label{fig:CN-sil}
\end{figure*}

\begin{figure}
        \centering
        \includegraphics[height=8.0cm,angle=-90]{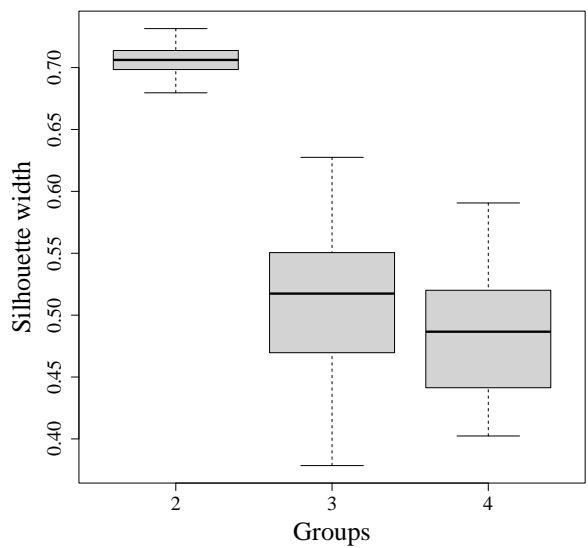}
        \caption{Boxplot of the average silhouette widths for two, three, and four groups obtained by MC resampling on the low-resolution spectroscopy data set.}
        \label{fig:CN-silMC}
\end{figure}

\subsection{Low-resolution spectroscopy}\label{sec:res-low}

Figure~\ref{fig:CN-sil} shows the results for the silhouette analysis applied to divisive HC on low-resolution spectroscopy data. This first result takes measurement errors only in the construction of the dissimilarity matrix into account. The left panel displays a two-group subsetting, while a three-group split is shown in the right panel. 
Different colours identify the subgroups. Objects are ordered from top to bottom in every group by decreasing silhouette value. The silhouette is bounded between -1 and 1; positive values indicate that the object fits its group well, and negative values suggesting that the object might be poorly classified. For every group, the size and average silhouette are shown (e.g. the upper group in the left panel is composed of 42 stars and has an average silhouette of 0.76).

The two-group division is highly favoured by the silhouette analysis as its average silhouette is 0.73. On the other hand, in the three-group case (right panel), the average silhouette is only 0.50. In particular, the three-group substructure is severely penalised by the fact that several stars assigned to the third group (green bars) fit poorly there, showing negative silhouette indexes.

\begin{table}[ht]
        \caption{Average silhouette values according to the clustering algorithm (by column) and number of proposed groups (by row).}
        \label{tab:CN-sil} 
        \centering
        \begin{tabular}{lccc}
                \hline
                Groups & Agglomerative HC & Divisive HC & PAM \\ 
                \hline
2 & 0.68 & 0.73 & 0.72 \\
3 & 0.60 & 0.50 & 0.62
\\
4 & 0.58 & 0.51 & 0.56 \\
        \hline
\end{tabular}
\end{table}

Table~\ref{tab:CN-sil} shows the average silhouette values obtained by the three clustering algorithms in the case of two, three, and four groups. All of them suggest that a two-group split is the optimum for the data set. However, it should be stressed that the maximum average silhouette we obtained, that is, 0.73, is quite low. As a rule of thumb, values above 0.7 are considered high enough to possibly originate from an actual population substructure and not from a random fluctuation \citep{Rousseeuw1987,KaufmanRousseeuw90}. The obtained value is just above this soft boundary, which means that the proposed MP identification is suspiciously weak and should be considered with great care.

\begin{figure}
        \centering
        \includegraphics[height=8.0cm,angle=-90]{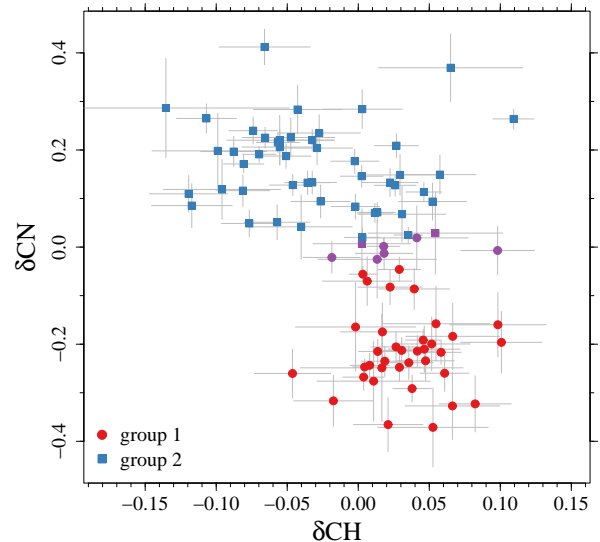}
        \caption{Scatter plot in the $\delta$CN vs $\delta$CH plane for the low-resolution spectroscopy data set. Data are classified according to the MC resampling. Circles correspond to objects classified into group 1, and squares show objects classified into group 2. Colours were adopted to distinguish data according to the cluster membership probability.  Purple symbols correspond to data for which the classification is uncertain (see text for details). }
        \label{fig:CN-scatter}
\end{figure}

As detailed in Sect.~\ref{sec:method}, we explored the effect of data uncertainty by repeating the cluster analysis and relying on MC data resampling. Figure~\ref{fig:CN-silMC} shows a boxplot of the average silhouette values for two, three, and four groups, obtained from 50 MC experiments. For each of them, the objects were perturbed according to a multivariate Gaussian distribution that accounted for the measurement errors.   
This result confirms that a two-group substructure is highly preferred.

Figure~\ref{fig:CN-scatter} shows the stars in the $\delta$CH - $\delta$CN plane classified according to the measurement error divisive HC algorithm. A slightly different approach was adopted in this case. Every star was perturbed $k = 21$ times, and all the $91 \times 21$ objects were retained in the data set. This full data set was then subjected to the HC analysis, as suggested by \citet{Su2018}. Then, for every object, the probability of falling into group $i$ was obtained by dividing the times $f_i$ it fell in group $i$ by $k$: $p_i = f_i/k$. The object was then assigned to the group corresponding to the maximum $p_i$. The two groups are identified by different symbols. The colours were adopted to code the information about the confidence in the grouping. The eight objects for which the maximum $p_i$ was lower than 0.75 are represented by purple symbols.
 
 \begin{figure}
        \centering
        \includegraphics[height=8cm,angle=-90]{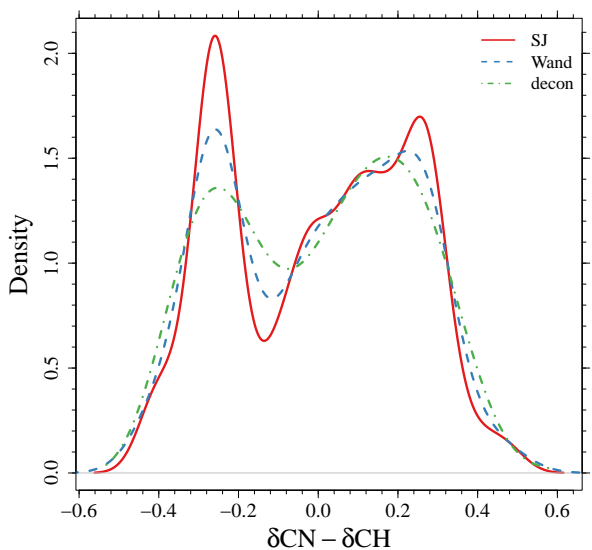}
        \caption{Kernel density estimators of the difference $\delta$CN~-~$\delta$CH, obtained by adopting the SJ bandwidth (solid red line), the Wand bandwidth (dashed blue line), and the approach from \citet{Wang2011} to correct for measurement errors (dot-dashed green line). }
        \label{fig:CN-density}
 \end{figure}
 
The analysis presented so far suggests a possible presence of two subpopulations from the low-resolution spectroscopy data set. However, relying on the analysis of the histogram of $\delta$CN~-~$\delta$CH, \citet{Hong2021} found four distinct populations. Thus, to explore the possible sources of these different results, we devote the final part of this section to study some potential issues arising from a histogram as a tool for detecting MP.

Figure~\ref{fig:CN-density} shows three different kernel density estimators obtained by adopting the SJ and Wand bandwidths. The values of the two bandwidths are 0.046 and 0.066 for SJ and Wand, respectively. The figure shows that the Wand bandwidth produces a smoother density estimator with two clear peaks. The adoption of the smaller SJ bandwidth results in a density with more features than the previous one, with the possible presence of a third small peak at $\delta$CN~-~$\delta$CH $\approx$ 0.1. This simple exercise shows that the adoption of an empirical density estimator for MP recognition is a quite dangerous approach and can lead to false-positive claims (see Sect.~\ref{sec:peaks} for further details). The adoption of the {\it decon} approach produces a density that looks more similar to that obtained with the Wand bandwidth than to that estimated using the SJ method.

The Dip test of multimodality showed highly significant evidence that the distribution is multimodal, producing a $p$ value of 0.01\footnote{The $p$ value is the probability of obtaining a test result as extreme as the actually observed result, assuming that the null hypothesis $H_0$ is true. Conventionally, a $p$ value lower than 0.05 is considered statistically significant and leads to the rejection of $H_0$.}. The result was confirmed by the excess mass test, which produced a $p$ value of 0.002. 

\begin{figure*}
        \centering
        \includegraphics[height=16.0cm,angle=-90]{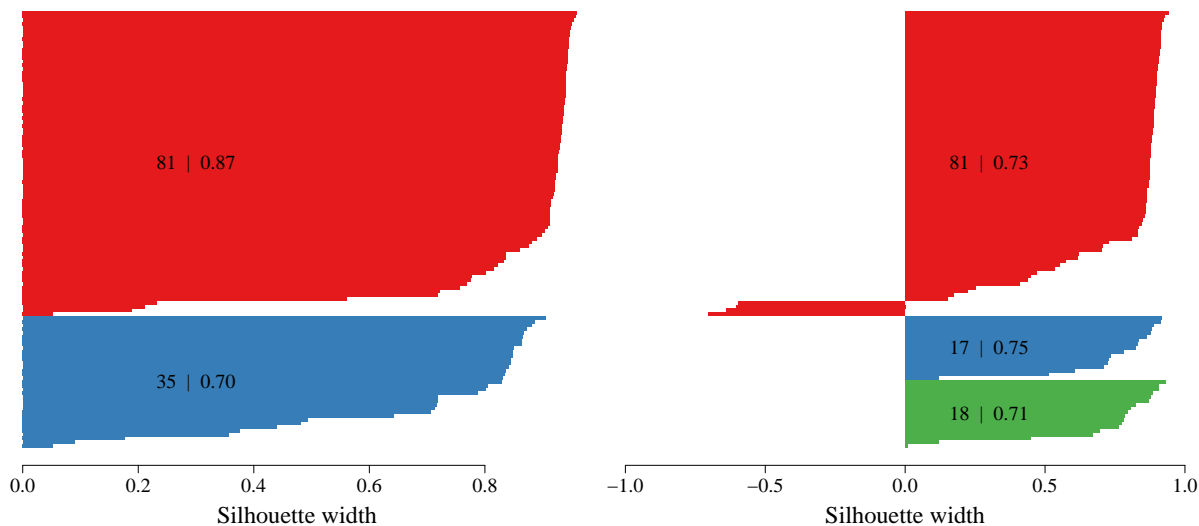}
        \caption{Same as in Fig.~\ref{fig:CN-sil}, but for the high-resolution spectrum data set.}
        \label{fig:Car-sil}
\end{figure*}

\begin{figure}
        \centering
        \includegraphics[height=8.0cm,angle=-90]{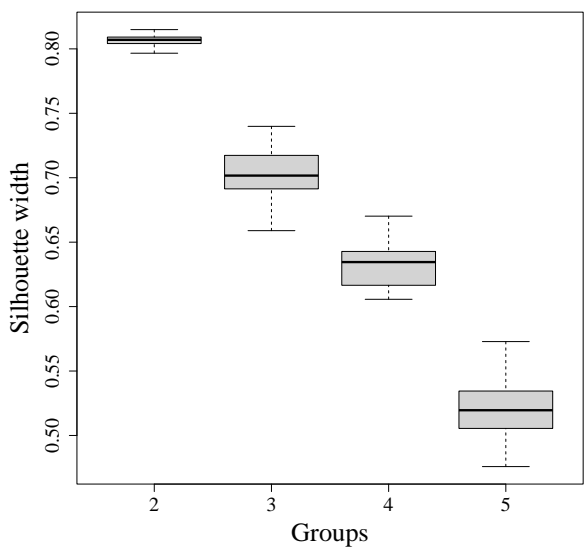}
        \caption{Boxplot of the average silhouette widths for two, three, four, and five groups obtained by MC resampling on the high-resolution spectroscopy data set.}
        \label{fig:Car-silMC}
\end{figure}

\subsection{High-resolution spectroscopy}\label{sec:res-high}

The analysis presented in Sect.~\ref{sec:res-low} was repeated on the high-resolution spectroscopy data set. Figure~\ref{fig:Car-sil} shows the silhouette plots for two and three subpopulations. A two-group split is preferred in this case as well, with an average silhouette width of 0.82 with respect to 0.73 for three groups. The value of the average silhouette width is higher than for low-resolution spectroscopy, which means that in this case, the detected substructure is more reliable.  

\begin{figure*}
        \centering
        \includegraphics[height=17.0cm,angle=-90]{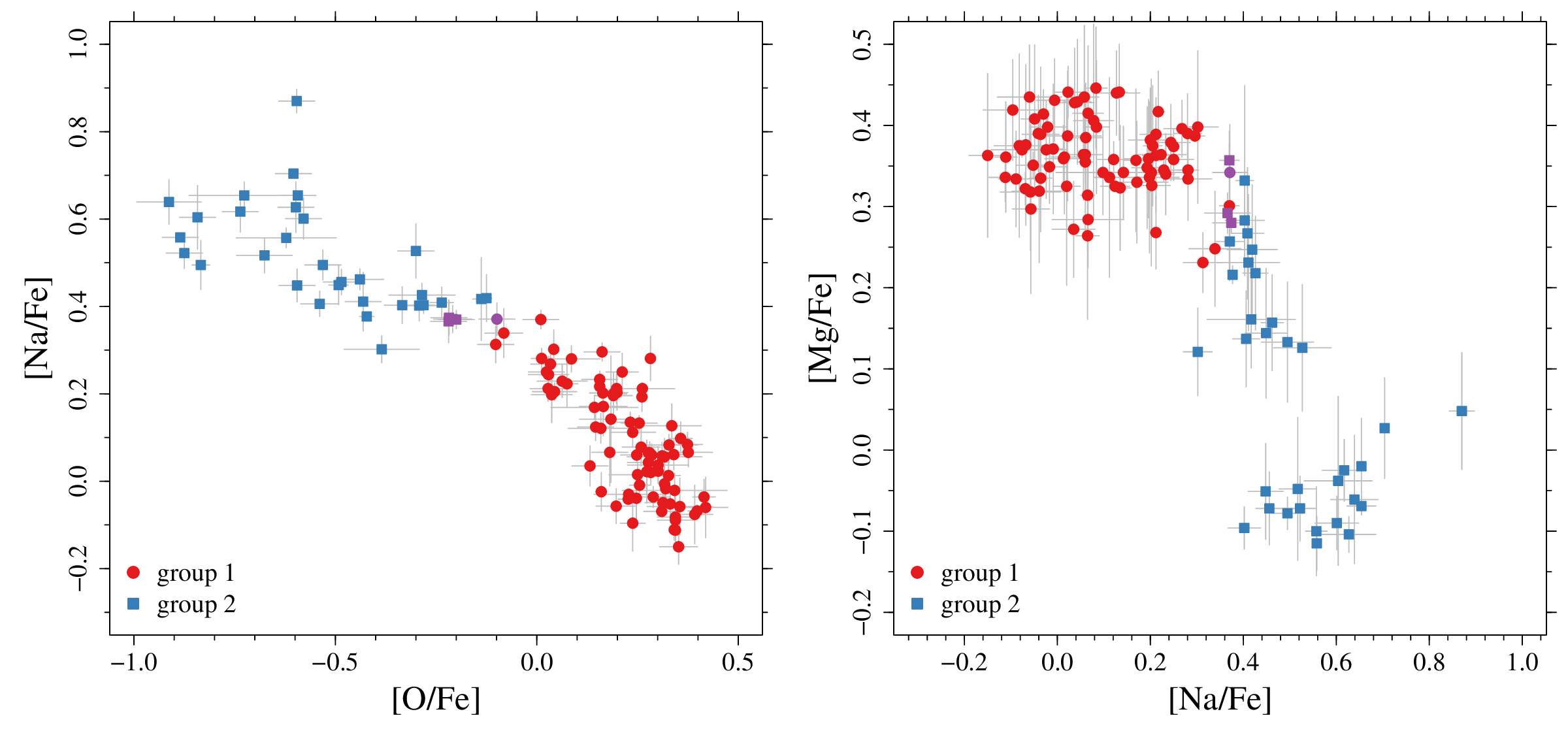}
        \caption{{\it Left}: Scatter plot in the [Na/Fe] vs [O/Fe] plane for the high-resolution spectroscopy data set. Data are classified according to the MC resampling. Circles correspond to objects classified into group 1, and squares show objects classified into group 2. Colours were adopted to distinguish data according to the cluster membership probability.  Purple symbols correspond to data for which the classification is uncertain (see text for details). {\it Right}: Same as in the left panel, but in the [Mg/Fe] vs [Na/Fe] plane.}
        \label{fig:Car-scatter}
\end{figure*}

The MC resampling was adopted to estimate the effect of measurement errors on the clustering. Figure~\ref{fig:Car-silMC} shows the boxplot of the average silhouette widths from two to five groups (the number of subpopulations suggested by \citealt{Carretta2015} in their analysis). The median value of the silhouette width decreases monotonically. Ultimately, even in the high-resolution spectroscopy, a two-group configuration is preferred for RGB stars in NGC 2808.

Figure~\ref{fig:Car-scatter} shows two scatter plots in the [Na/Fe] versus [O/Fe] and [Mg/Fe] versus [Na/Fe] planes. The object memberships were established by the same MC procedure as described for the low-resolution spectroscopy data set. Only four objects have a questionable membership and are shown in purple in the figure. 

\begin{figure}
        \centering
        \includegraphics[height=8.0cm,angle=-90]{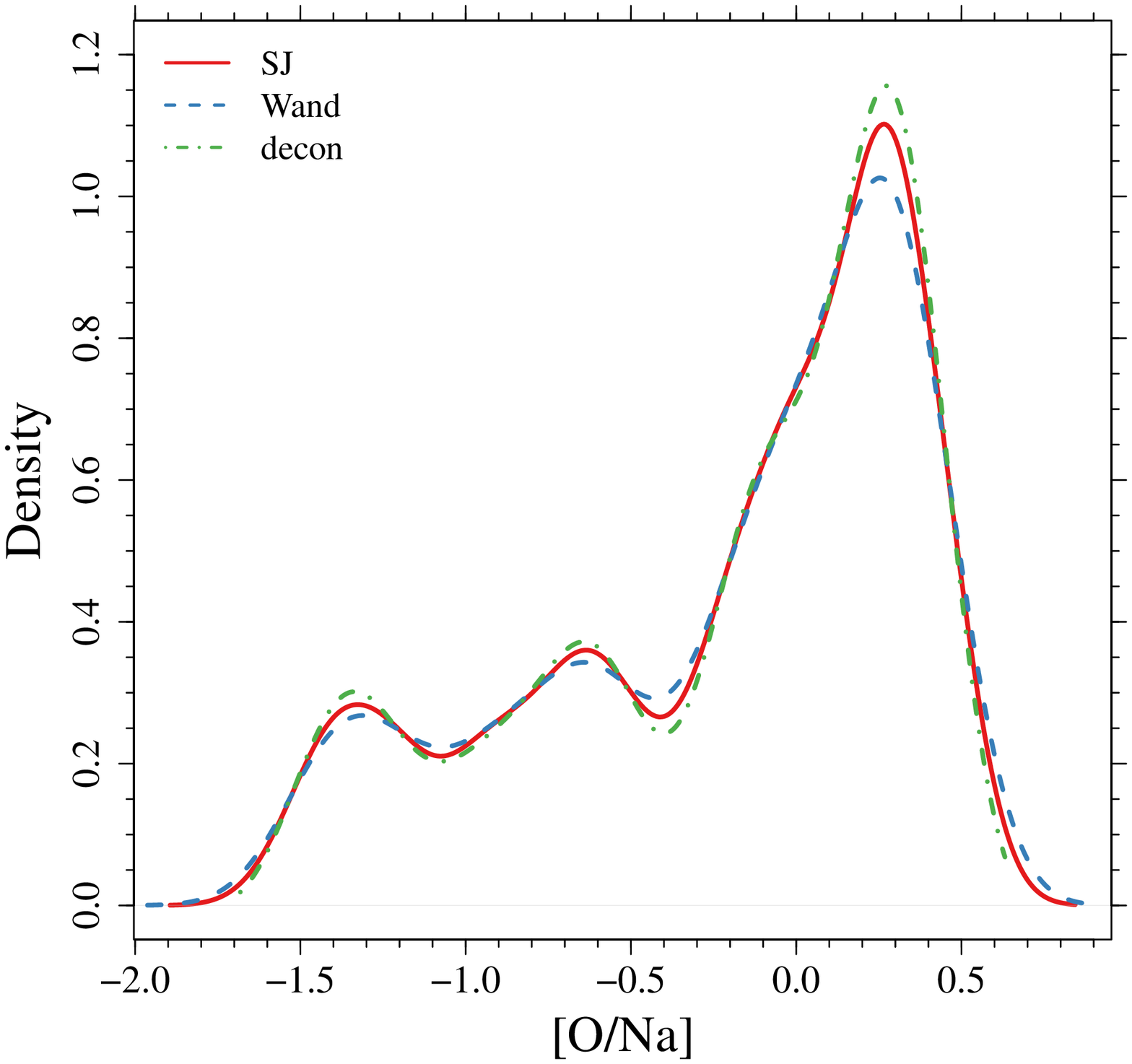}
        \caption{Kernel density estimators of [O/Na], obtained adopting the SJ bandwidth (solid red line), the Wand bandwidth (dashed blue line), and the approach from \citet{Wang2011} to correct for measurement errors (dot-dashed green line).}
        \label{fig:Car-density}
\end{figure}

Finally, Fig.~\ref{fig:Car-density} shows three kernel density estimators of the [O/Na] probability density function. The abundance ratio [O/Na] was used by \citet{Carretta2015} to propose the presence of five MPs by the peaks detected in the histogram. The three methods we adopted give very similar results: a strong peak at [O/Na] $\approx$ 0.3 and two much smaller peaks around $-0.6$ and $-1.4$. The difference in the peak heights should be carefully considered because it means that only few data are in the low [O/Na] zone, so that fluctuations are much more probable there. The Dip test $p$ value for the presence of more than one peak was 0.94, therefore we cannot reject the hypothesis of a unimodal distribution. The conclusion was confirmed by the excess mass test ($p = 0.58$).
Ultimately, as already mentioned, the peak count cannot be considered a reliable tool for MP identification. We discuss this issue in more detail in the next section.
    
\subsection{Issues in identifying MP by peaks in the histograms}\label{sec:peaks}    
\begin{figure*}
        \centering
        \includegraphics[height=16.0cm,angle=-90]{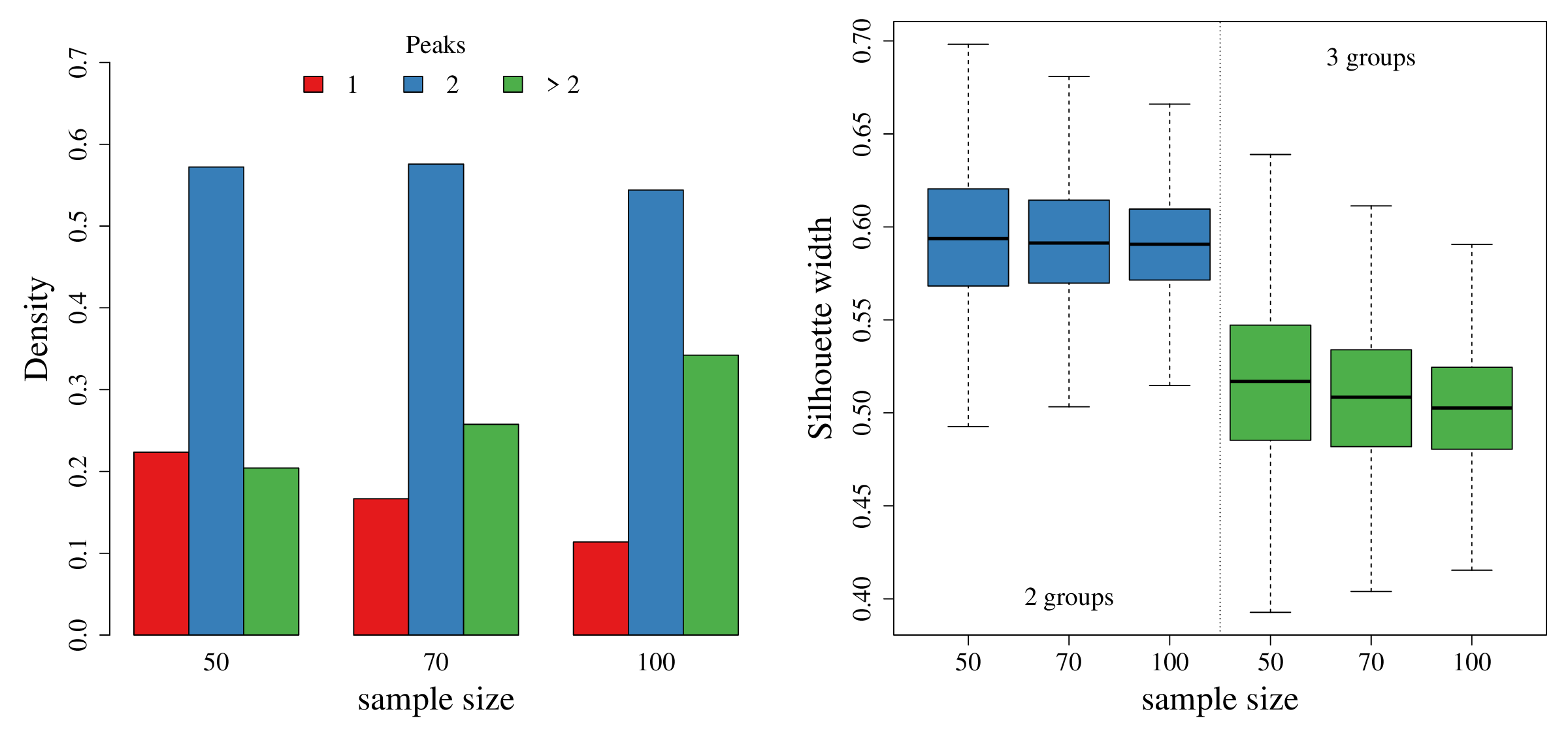}
        \caption{{\it Left}: Empirical density function for the number of peaks detected in the MC simulation, classified according to the sample size (see text). {\it Right}:  Boxplots of the average silhouette widths for two- and three-group splits, according to the sample size.  }
        \label{fig:peaks}
\end{figure*}
 
 To show some issues arising from the adopting an empirical density estimator such as a kernel density or a histogram, we present here a simple exercise. We simulated random points from a single population with an arc-shaped distribution in 2D space to mimic the abundance plots in Figs.~\ref{fig:CN-scatter} and \ref{fig:Car-scatter}, which we call $P_i = \{x_i, y_i\}$. As a simple choice, we sampled the points from a circle with a unit radius and angular coordinate $\theta \in [0, \pi/2]$, adopting a uniform distribution for $\theta$. The points were then perturbed with Gaussian errors with $\sigma = 0.06$ in both dimensions. Then, we computed the kernel density estimator of the coordinate difference $z_i = x_i - y_i$, similarly to the process of inferring peaks in the element abundances carried out in \citet{Carretta2015} and \citet{Hong2021}. The SJ bandwidth estimator was adopted. The peaks in the kernel density estimator were identified by the change in the sign of the first derivative. Finally, we also performed a divisive HC analysis and computed the average silhouette widths for two and three subgroups. The process was repeated 5,000 times for three different population sizes: 50, 70, and 100 objects.

The results we obtained by adopting a uniform distribution for the angular coordinate $\theta$ are presented in Fig.~\ref{fig:peaks}. The left panel shows that the detection of multiple peaks is indeed very probable in this scenario, with two peaks being the most frequent finding. As the sample size increases, the detection of two or more peaks becomes increasingly probable because the higher the number of points, the higher the occurrence of mini clusters, which in turn leads to peaks in the density estimator. The right panel of the figure shows the boxplots of the average silhouette widths for the two- and three-group split. The median of these boxplots is reassuringly low, it is about 0.60 for two groups and about 0.50 for three groups. Therefore the cluster analysis leads to the rejection of the hypothesis of the existence of MPs in the parent population.      
Multiple peaks can even be expected and are partially due to the way in which the synthetic data were simulated. The adoption of a flat sampling enhances the possibility of a random aggregation of a few points, thus leading to peaks in the density function. Nevertheless, we remark that the peaks identified in the density estimator analysis are rejected by the silhouette plots, showing the robustness of this second method.
  
In the real world, the assumptions of this exercise may not hold perfectly because the scatter plot of the element abundance may differ from a perfect arc, and the angular distribution may not be uniform. To verify the result robustness, we repeated the exercise and adopted a single-peak angular distribution: a Gaussian angular distribution peaked at $\pi/4$ and standard deviation $\pi/12$. Even in this case, we obtained a relevant probability of a multiple peak ($\geq 2$) detection using a kernel density analysis, from 30\% for a sample size of 50 objects to 22\% for 100 objects. On the other hand, as in the previous case, the silhouette analysis rejected the presence of multiple groups in the synthetic data set. 

In summary, the results presented in this section are very relevant when the number of subgroups is judged by counting peaks in the empirical density. They should be considered as an alarm bell showing that empirical density estimators are prone to peak detection even when no peak is present.

\section{Discussion and conclusions}\label{sec:conclusions}

This paper investigated the adoption of robust statistical clustering methods to assess the presence of MPs in a globular cluster by considering the abundances of some key elements alone.
We explicitly addressed the problem of clustering data in the presence of measurement errors by adopting techniques that were recently proposed in the statistical literature \citep[e.g.][]{Kumar2007, Su2018}. 

The globular cluster NGC 2808 was chosen as case study because it hosts several MP reported in the literature, from the main sequence to the asymptotic giant phase  \citep[see e.g.][]{DAntona2005, Piotto2007, DAntona2008, Lee2009, Dalessandro2011, Carretta2015, Marino2017}. In particular, we focused our analysis on the RGB evolutionary phase, where two high-quality data sets are available:  \citet{Carretta2015} for high-resolution spectroscopy and \citet{Hong2021} for low-resolution spectroscopy.

Several statistical clustering techniques were applied to the two data sets. Two HC approaches (divisive and agglomerative) and a partition method (partition around medoids) were adopted. As they gave similar results on both data sets, we only discussed those obtained by the divisive HC algorithm. The dendrograms resulting from the HC algorithm were analysed by means of the silhouette plot method to compare the performances of different numbers of subgroups \citep{Rousseeuw1987, KaufmanRousseeuw90}. This is a general-purpose well-established method for investigating the structure of a dendrogram; adopting this has the fundamental merit of making the population split objective and reproducible among different investigations.

For both data sets, the analysis suggested a possible split into two subpopulations as optimal. The claim is stronger for the high-resolution spectroscopy data, with a 0.82 average silhouette width for a two-group division. In contrast, the average silhouette width for two groups from low-resolution spectroscopy was only 0.73. As a rule of thumb, values higher than 0.7 are considered due to an actual population substructure and not to a mere fluctuation. 

The finding of only two subpopulations in NGB 2808 is at odds with the five subgroups reported by \citet{Carretta2015} for high-resolution spectroscopy and the four MPs by \citet{Hong2021} for low-resolution spectroscopy. These differences are easily due to the difference in the methods adopted to define the presence of a distinct populations. In the two quoted papers, the estimated number of MPs relied on the analysis of the empirical distribution of some key abundances by means of histograms. Peaks were considered linked to the presence of a possible subpopulation, and these findings were corroborated by analysing the scatter plot of different elements to confirm by naked eye that the proposed division holds. Several shortcomings make this approach suboptimal. First, it only adopts part of the data to define the peaks and uses the remaining part as a confirmation. A multivariate approach to classification problems has the advantage of considering all the data at once, offering a much more robust classification. This was recognised  by \citet{Fisher1936} in his seminal paper about multivariate analysis. Second, as we directly showed here with an MC experiment, histogram analysis is prone to false-peak detection. Sampling data uniformly from an arc-shaped distribution, we showed that multiple peaks are detected by empirical distribution estimators in about 85\% of the experiments. We also showed that the divisive HC method, followed by silhouette width analysis, did not suggest an underlying substructure in any population. Third, the presence of peaks was established by subjective judgement and not by sound statistical methods. The adoption of the Dip test and of the mass excess test \citep{Hartigan1985, Ameijeiras2019} to verify the presence of multimodality provided significant departure from unimodality for the low-resolution spectroscopy data set, but no evidence against the hypothesis of a single mode in the high-resolution case.

Particular emphasis  was given to the correction needed for classical clustering algorithms in the presence of measurement errors. In this paper we adopted the approach by \citet{Su2018}, based on MC resampling and data perturbation. In the presence of data uncertainties, clustering methods cannot provide a firm membership for the various objects, but the probability that they fall into the different groups. Objects can then be assigned to the group with the highest probability. We found that for both data sets, the vast majority of stars were clearly classified into a group (membership probability higher than 75\%); only eight stars from low-resolution spectroscopy and four from high-resolution spectroscopy had a questionable membership.     

The analysis presented in this paper only used element abundances to define the population substructure. However, photometric pseudo-colours are often adopted as a proxy to trace variation in the stellar abundances. Several pseudo-indexes were proposed in the literature that are sensitive to different element variations \citep[see e.g.][]{Marino2008, Milone2015, Milone2017, Bastian2018}. An advantage of the techniques adopted in our analysis is that they can be easily extended to incorporate this information whenever available. A mixed data set containing both abundances and colour indexes requires some care because the ranges they span can be somewhat different. In this case, it is advised to standardise the data (and scale the associated uncertainties) to zero mean and unit variance before the dissimilarity matrix is computed.

The increasing availability of high-quality data facilitates the analysis of the presence of MPs in globular cluster more than ever. It is therefore of paramount importance to base  these analyses  on firm statistical ground and not on naked-eye assessment.  
While we presented  methods that are well established in the statistical literature, many other exist and the research in the field is ongoing: as an example, we can mention clustering by Gaussian mixture models, density-based spatial clustering of applications with noise
\citep{Ester1996}, or clustering in generative adversarial networks \citep[e.g.][]{He2020, Pal2021}. 
Ultimately, the adoption of methods analogous to those discussed here can help to obtain more robust and reproducible researches.

\begin{acknowledgements}
  The authors acknowledge the anonymous referee for the useful comments that helped to improve the clarity of the paper. ET is funded by the Czech Science Foundation GA\u{C}R (Project: 21-16583M). He also thanks Pisa University for the hospitality during 6/09/2021 - 5/10/2021 through the visiting program.
\end{acknowledgements}

\bibliographystyle{aa}
\bibliography{biblio}

\appendix

\section{Statistical tools for data clustering}
\label{sec:app-tools}

Let $\mathbf X$ be the a $n \times p$ matrix of the $p$ observed quantities for the $n$ objects under consideration. Let $x_{ij}$ be the element of the $i$-th row and $j$-th column of $\mathbf X$, and  $\sigma_{ij}$ its uncertainty. We aim to split the $n$ objects into a natural grouping, taking the measurement errors into account. We can define the \emph{dissimilarity matrix} $\mathbf D$ for the $n$ objects as the matrix whose elements $d(a,b)$ are the squared Euclidean distances between rows $a$ and $b$ of $\mathbf X$, weighted by their errors, namely,
\begin{equation}
        d(a,b) = \sum_{j=1}^p \left(\frac{x_{aj} - x_{bj}}{\sqrt{\sigma_{aj}^2 + \sigma_{bj}^2}} \right)^2 
        \label{eq:dissim}
.\end{equation}
The adoption of the squared distances helps to better show substructures in the data. This produces a much clearer separation.

In the following we also use a distance among clusters, defined  as follows. Let $A$ and $B$ be two objects joined in a single group $A + B$. The distance between this group and a group $C$ is evaluated using the following expression:
\begin{eqnarray}
        d(C, A+B) & = & \delta_1 d(C,A) + \delta_2 d(C,B) + \delta_3 d(A,B) + \nonumber\\
        & + &  \delta_4[d(C,A) - d(C,B)], \label{eq:distance}
\end{eqnarray}
where $d(C, A)$, $d(C, B)$, and $d(A, B)$ are computed using Eq.~(\ref{eq:dissim}), and $\delta_{1, \ldots, 4}$ are some weight factors that have to be defined depending on the purpose of the analysis (see the next appendix).

The cluster analysis computations described in the following sections were performed using R 4.1.0 \citep{R} by means of the functions in  the package {\it cluster} \citep{cluster2021}.

\subsection{Agglomerative hierarchical clustering}
\label{sec:agglom}

The technique starts with each observation forming a cluster by itself. At each step, the clustering algorithm  merges the two nearest clusters. After the first step, the two nearest objects are merged. For the second step, we used the distance among clusters defined by eq.(\ref{eq:distance}). In this equation, we used  the Ward weighting for $\delta_{1, \ldots, 4}$, a choice that minimises the heterogeneity within clusters \citep[see e.g.][]{simar}, namely
\begin{eqnarray}
        \delta_1 & = & \frac{n_C+n_A}{n_A+n_B+n_C}\nonumber\\
        \delta_2 & = & \frac{n_C+n_B}{n_A+n_B+n_C}\nonumber\\
        \delta_3 & = & -\frac{n_C}{n_A+n_B+n_C}\nonumber\\
        \delta_4 & = & 0.
\end{eqnarray}
The clustering was repeated until all the observations were in the same cluster.

\subsection{Divisive hierarchical clustering}
\label{sec:divis}

The technique starts with all observations grouped into a cluster. At each step, the cluster with the largest dissimilarity  $d(a,b)$ between any two of its members was split into two subgroups. The element with the largest average dissimilarity to the other observations of the selected cluster was chosen as the new group progenitor, and the method reassigns observations considering whether their average dissimilarity to the new group members is lower than to the members of the original group. Therefore only the matrix $D$ is needed for divisive HC. As for agglomerative HC, the divisive approach produces a dendrogram.

\begin{figure*}
        \centering
        \includegraphics[height=16.0cm,angle=-90]{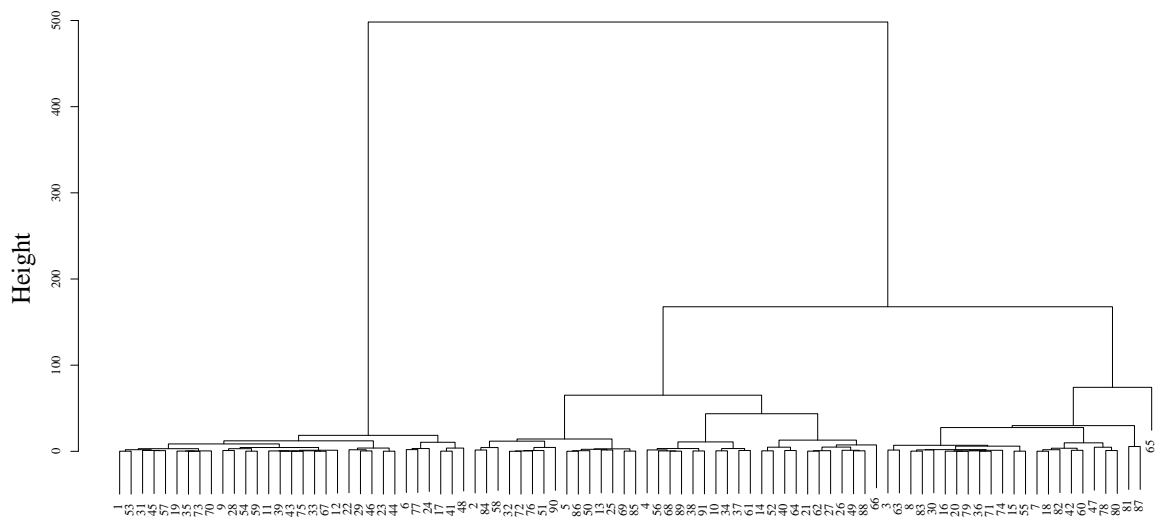}
        \caption{Dendrogram obtained by a divisive HC on the low-resolution spectroscopy data set. The leafs are labelled by the data set row in which the object lies. The lower two objects merge, the higher their similarity. Cutting the tree at different heights yields a different number of subgroups. }
        \label{fig:CN-dendro}
\end{figure*}

The result of an HC analysis is a dendrogram, as shown in Fig.~\ref{fig:CN-dendro} obtained with the low-resolution spectroscopy data set. Leafs are identified by the row number of the corresponding object in the data set. 
The height of the nodes is the distance, as
defined in Eq.~(\ref{eq:distance}), at which the corresponding clusters merge.
The lower a node, the more similar the merged clusters.
Cutting the dendrogram at different heights 
produces a different number of subgroups. 
The optimal number of groups suggested by the clustering was determined
according to the silhouette plot analysis (see Appendix~\ref{sec:sil}).

A known problem of HC methods (both agglomerative or divisive) is the fact that when an object is joined to a cluster, it will not be considered again, so that the method cannot reconsider its steps to possibly provide a better clustering. This problem is more severe for agglomerative HC, which merges neighbour points without considering the global distribution of data. These decisions cannot be reconsidered. In contrast, divisive HC takes the global dissimilarity of data into account when partitioning decisions are made. Another problem is that all HC techniques are sensitive to outliers. Finally, but this is not relevant here, the computational complexity of agglomerative HC is $O(n^2)$, which means that it can only be used in small data sets \citep{KaufmanRousseeuw90,SAXENA2017}.   

\subsection{Silhouette analysis}
\label{sec:sil}

The result of HC is a dendrogram, which does not carry any information about the optimal split of data. To determine how well different partitioning works, the dendrogram can be subjected a further analysis, evaluating the performance of different subgroupings. Several indexes have been developed to this purpose, none of which is clearly optimal. However, the average silhouette has proven effective in most circumstances  \citep{Bezdek1998, DEAMORIM2015}.

This method provides a value, called silhouette, that measures how well an object lies within its cluster. The clustering providing the largest average silhouette is chosen  as the best \citep{Rousseeuw1987, KaufmanRousseeuw90}.

The algorithmic approach to constructing a silhouette is the following. For each
object $i,$ let $\bar d(i)$ be the average dissimilarity among $i$ and all other objects
inside the same cluster, computed using Eq.~(\ref{eq:dissim}). Then, for all the other clusters $C$, let $\bar d(i,C)$ be the average dissimilarity of $i$ and all the elements of $C$ (again computed using Eq.~(\ref{eq:dissim})). Let $b(i)$ be the
minimum value of $\bar d(i,C)$ over all the clusters $C$. The silhouette value $s(i)$ is
then evaluated as
\begin{equation}  
        s(i) = \frac{b(i)-\bar d(i)}{{\rm max}\{\bar d(i), b(i)\}}.
\end{equation}

When a cluster contains a single object, then by definition $s(i) = 0$. A high value of $s(i)$ implies that an observation lies well inside its cluster, while a value near 0 indicates that the observation lies equally well inside its cluster or in the nearest cluster. A negative silhouette suggests that the object might be in the wrong group. By averaging the values of $s(i)$ over all the objects, the average silhouette is obtained that is used for diagnostic purposes. Usually, the clustering is considered reliable when the average silhouette is higher than 0.7, suspiciously spurious for average silhouette between 0.5 and 0.7, and unreliable for average silhouette below 0.5 \citep{Rousseeuw1987,KaufmanRousseeuw90}.

\subsection{Cluster analysis in the presence of measurement errors}
\label{sec:error-var}

Classical cluster analysis methods ignore the uncertainty associated with data.
Recent theoretic developments addressed the measurement error
problems in the framework of unsupervised clustering \citep[see e.g.][]{Kumar2007, Su2018}. 
In this paper we adopted the approach proposed by \citet{Su2018}, which relies on MC resampling and has been proven to be capable of reproducing the clusters that would have been formed if the variable had been observed without errors.

The method consists of generating for each row of $\mathbf X$ ($x_i$) $J$ MC synthetic observations, drawing them from the asymptotic error distribution. We adopted a multivariate Gaussian error distribution with mean $\mathbf 0$ and covariance matrix $\mathbf \Sigma = {\rm diag}(\sigma_i^2)$. After this generation, the data set contained $n \times J$ observations. This large data set was then subjected to the cluster analysis with any classic clustering method. Due to its probabilistic nature, the algorithm does not provide a firm classification of the parent objects; its result is the probability that a given object is contained in a given cluster. We adopted the algorithm with $J=1$, repeating the MC simulation 50 times. This approach allowed us to estimate the variability of the silhouette index given the uncertainties in the data. 

\subsection{Partitioning methods}
\label{sec:pam}
Partitioning methods such as the K-means clustering algorithm or partition around medoids (PAM) choose a prespecified number of cluster centres to minimise the within-class sum of squared distances from these centres. They are most appropriate to discover hyper-spherical clusters around these centres \citep{venables2002modern, simar}. 
When the centroids are confined to objects in the data set, the method is known as medoids criterion, resulting in an approach that is more robust to outliers.

The algorithm a priori assumes the number $k$ of groups, then 
selects $k$ points from the data set to minimize the intercluster squared distances. 
The best number of groups can then be assessed a posteriori by the analysis of the cluster efficiency, for instance, by means of silhouette analysis.
As the splitting problem is NP-hard, many heuristic solutions exist. After the first step, each point is associated with the closed medoid (in terms of the specified distance). Then an iterative procedure attempts to swap for every cluster the medoid with every given point inside the cluster. If a new medoid gives a lower sum of squared distances, it is retained as the new medoid, and the membership of all other points is reassessed. Otherwise, the algorithm terminates \citep{KaufmanRousseeuw90}.

\subsection{Kernel density estimator}
\label{sec:kernel}

The kernel density is a non-parametric estimate of the probability density
function from a set of data. It is a generalisation of
the histogram, with better theoretical properties \citep{simar}. For a set of
$n$ observations $x_1$, $x_2$, $\ldots$, $x_n,$ a kernel density with bandwidth
$h$ has the form
\begin{equation}    
        \hat f(x,h) = \frac{1}{n h} \sum_{i=1}^n K\left(\frac{x-x_i}{h}\right)
,\end{equation}    
where the kernel function $K$ is chosen to be a probability density function.
Several choices of kernel are available. We used a
Gaussian kernel,
\begin{equation}    
        K(y) = \frac{1}{\sqrt{2 \pi}} \exp\left( - \frac{y^2}{2}\right).
\end{equation}
The kernel selection usually affects the kernel estimate with
respect to the bandwidth $h$ only little. This parameter is selected by balancing two effects
because an increment of $h$ increases the bias of $\hat f$ while it reduces its
variance. Several choices for the bandwidth, based on the asymptotic expansion of the mean
integrated squared error, are reported in the literature. The different
choices have an effect on the kernel estimator for multimodal
distributions ( \citet{Feigelson2012, simar, venables2002modern, Sheathe1991}).
Because of their excellent properties, we adopted the \citet{Sheathe1991} and \citet{Wand1994} plug-in bandwidths as reference. 

The computation of the kernel density estimator when data are measured with known uncertainty was addressed extensively in past decades. \citet{Carroll1988} proposed the deconvolution kernel density estimator to recover the unknown density function from contaminated data. Since then, the problem received great attention \citep[see e.g.][]{Zhang1990, Efromovich1997, Delaigle2004}. Recently, the problem of heteroscedastic errors was also addressed \citep[see among others][]{Wang2011, Achilleos2012}. We adopted this method as implemented in the R library {\it decon} \citep{Wang2011}.

\end{document}